\documentclass[]{aa}
\input{epsf}

\begin{document}
\newcommand{\hubble}{\mbox{\,km\,sec$^{-1}$\,Mpc$^{-1}$}}
\def \map{\hbox{$\langle M_{\rm LSS}^2\rangle^{1/2}$}}

\title{The effect of distant large scale structure on weak lensing
mass estimates}

\author{Henk Hoekstra$^1$}

%\offprints{H.~Hoekstra}

\institute{Kapteyn Astronomical Institute, University of Groningen,
P.O.~Box 800, 9700 AV Groningen, The Netherlands\\
email: hoekstra@astro.rug.nl}

   \date{Received -; accepted -}

\authorrunning{H. Hoekstra}
\titlerunning{Effect of LSS on mass estimates}

   \maketitle

\begin{abstract} We quantify the uncertainty in weak lensing mass
estimates of clusters of galaxies, caused by distant (uncorrelated)
large scale structure along the line of sight. We find that the effect
is fairly small for deep observations $(20<R<26)$ of massive clusters
($\sigma=1000$ km/s) at intermediate redshifts, where the bulk of the
sources are at high redshifts compared to the cluster redshift. If the
lensing signal is measured out to $1.5~h_{50}^{-1}$~Mpc the typical $1
\sigma$ relative uncertainty in the mass is about $6\%$. However, in
other situations the induced uncertainty can be larger. For instance,
in the case of nearby clusters, such as the Coma cluster, background
structures introduce a considerable uncertainty in the mass, limiting
the maximum achievable signal-to-noise ratio to $\sim 7$, even for
deep observations. The noise in the cluster mass estimate caused by
the large scale structure increases with increasing aperture size,
which will also complicate attempts to constrain cluster mass profiles
at large distances from the cluster centre. However, the distant large
scale structure studied here can be considered an additional
(statistical) source of error, and by averaging the results of several
clusters the noise is decreased.


   \end{abstract}

\section{Introduction}
\footnotetext[1]{Present address: CITA, University of Toronto, 60 St.
George Street, M5S 3H8 Ontario, Canada}

Light rays are deflected as they travel through the universe, and as a
result our view of the distant, faint galaxies is somewhat distorted.
Measuring these lensing induced distortions has proven to be a
valuable tool in observational cosmology (see the review by Mellier
1999): the lensing signal is sensitive to the projected mass surface
density of the deflector, and through lensing one can probe the mass
distribution without having to rely on assumptions about the dynamical
state and nature of the deflecting matter.

When the surface density of the lens is sufficiently high, the
distortions can be so large that multiple images can be seen, which
provide strong constraints on the mass distribution in cluster cores.
In this paper we will concentrate on the weak lensing regime, which
allows us to study the cluster mass distribution on larger scales.

Weak lensing is a powerful tool to probe the distribution of matter in the
universe on various scales. One exciting application is the
measurement of the lensing signal induced by large scale structure
(e.g.  Blandford et al. 1991; Kaiser 1992; Bernardeau, van Waerbeke,
\& Mellier 1997; Schneider et al. 1998, SWJK hereafter), and recently
several groups have reported detections of this signal (Bacon et
al. 2000; Kaiser, Wilson, \& Luppino 2000; van Waerbeke et al. 2000;
Wittman et al. 2000).

The first successful weak lensing results were obtained by
studying the systematic distortion of the images of faint galaxies
behind rich clusters of galaxies (e.g. Tyson, Wenk, \& Valdes 1990).
Measuring the masses and mass-to-light ratios of rich clusters is 
an important application of weak lensing, because comparison of the
weak lensing result with other mass estimators (e.g., X-ray, S-Z effect,
or velocities) can give important clues about the geometry and dynamical
state of the system (e.g, Zaroubi et al. 1998).

Weak lensing also has some drawbacks. First of all, the amplitude of
the lensing signal depends on the redshift distribution of the faint
background galaxies. These galaxies are generally too faint to be
included in spectroscopic redshift surveys, but photometric redshifts
are thought to provide ample information on the redshift distribution
of these galaxies (e.g., Hoekstra, Franx, \& Kuijken 2000). Also the
observed weak lensing signal is a measure of a surface density
contrast, and one of the problems associated with this is the mass
sheet degeneracy (Gorenstein, Shapiro, \& Falco 1988). One could try
to get around this problem by measuring the lensing signal out to
large distances from the centre, and arguing that the cluster surface
density should vanish there.

Another potential problem is the fact that the lensing
signal is sensitive to all matter along the line of sight.
In general, weak lensing analyses of clusters assume that the signal
induced by the cluster under investigation is so dominant that 
the contribution from other structures along the line of sight can be
neglected. 

The effect of structures, close to the cluster, such as the filaments
that connect the cluster to the `cosmic web', or an elongation along
the line of sight, have been studied through N-body simulations (e.g,
Cen 1997, Metzler et al. 1999). Local structures introduce both a bias
and an additional uncertainty in the cluster mass estimate, in
particular when one for instance tries to measure the mass within
$r_{200}$.

In this paper we study how distant structures affect weak lensing mass
estimates of clusters of galaxies. Unlike the local structure, these
are uncorrelated with the cluster, and they can be filaments, galaxy
groups, or even other clusters. That such structures affect the mass
estimates is easily seen in the case of a massive high redshift
cluster, with a low redshift group in front of it. Assume the cluster
is at a redshift of unity, with a group at say $z=0.3$, and that the
mass of the group is one tenth of that of the cluster: the
contribution of the group to the total lensing signal is about one third, 
compared to the 10\% in mass.

In this paper we will quantify the uncertainty in weak lensing mass
estimates of rich clusters caused by intervening matter. To do so we
examine the predicted statistical properties of the contribution by
large scale structure. We use an approach similar to what is done to
predict the amplitude of the cosmic shear signal, which is used to
constrain the cosmological parameters. Given a cosmological model,
we estimate the uncertainty in cluster masses derived from weak lensing
analyses.

The structure of this paper is as follows. In section~2 we discuss two
popular methods to measure cluster masses. In section~3 we derive the
equations that describe the effect of large scale structure on such
mass estimates.  The effect of large scale structure on mass estimates
for clusters is quantified in section~4. We examine the dependence
on cluster redshift, aperture size and source redshift distribution.

\section{Estimating masses from weak lensing}

Cluster masses can be measured in several ways. In this paper
we will use aperture mass statistics, which provide a filtered
measurement of the of the surface mass density in an aperture
(Schneider 1996, S96 hereafter):

\begin{equation}
M_{ap}(\theta)=\int_{0}^{\theta} d^2\theta' U(|\theta'|)\kappa(\theta'),
\end{equation}

\noindent where $U(\theta)$ is the weight or filter function. Provided
that $U(\theta)$ is compensated, i.e.

\begin{equation}
\int_0^{\theta} db~b U(b)=0,
\end{equation}

\noindent the aperture statistic can be related to the tangential
shear $\gamma_T$. Unfortunately $\gamma_T$ cannot be observed
directly, but one can measure the distortion
$g_T=\gamma_T/(1-\kappa)$.  In the weak lensing regime, where
$\kappa\approx 0$, one can use $\gamma_T\approx g_T$, which is what we
do throughout this paper.  Then we have

\begin{equation}
M_{ap}(\theta)=\int_0^{\theta} d^2\theta Q(|\theta|)\gamma_T(\theta)
\label{eqmapq}.
\end{equation}

\noindent Here, $Q(\theta)$ is related to $U(\theta)$ through:

\begin{equation}
Q(b)=\frac{2}{b^2}\int_0^{b} db' b' U(b') - U(b).
\end{equation}

S96 has shown that the signal-to-noise ratio of the aperture mass
measurement is optimal when $Q(|\theta|)$ is the same as the
(expected) tangential shear as a function of $|\theta|$.

We want to quantify the uncertainty in mass measurements of clusters,
and therefore require that the aperture statistic has a clear
physical meaning. Studies of cosmic shear (see SWJK) use
different choices for $Q(|\theta|)$. Below we discuss two statistics
that have been used to estimate cluster masses.

\subsection{SIS model fit}

A popular model to describe the cluster mass distribution is the
singular isothermal sphere (SIS)

\begin{equation}
\kappa(r)=\frac{r_E}{2r},
\end{equation}

\noindent where $r_E$ is the Einstein radius. The Einstein radius 
(in radians) is related to the velocity dispersion through

\begin{equation}
r_E=4\pi\left(\frac{\sigma}{c}\right)^2 \beta.
\end{equation}

\noindent The tangential shear as a function of radius is simply
given by

\begin{equation}
\gamma_T(r)=\kappa(r)=\frac{r_E}{2r}.
\end{equation}

\noindent The best fit SIS model is obtained by minimizing

\begin{equation}
\sum_i \left(\frac{\gamma_{T,i}-r_E/(2r_i)}{\sigma_\gamma}\right)^2,
\end{equation}

\noindent with respect to $r_E$, where $\gamma_{T,i}$ are the
individual measurements of the lensing signal at radius $r_i$. The
uncertainty in the measurement of the tangential shear,
$\sigma_\gamma$ is given by

\begin{equation}
\sigma^2_\gamma=\frac{\sigma^2_{\rm gal}}{2\pi\bar n r_i \Delta r_i},
\end{equation}

\noindent with $\sigma_{\rm gal}$ the uncertainty in the measurement
of an individual galaxy, $\bar n$ the number density of sources, and
$\Delta r_i$ the width of the $i$th radial bin. If we take the widths
of all the bins the same, and assume that the number density of sources
is uniform, minimizing with respect to $r_E$ gives

\begin{equation}
r_E=2 \frac{\sum_i \gamma_{T,i}}{\sum_i\frac{1}{r_i}}~\label{eqre}.
\end{equation}

For a continuous profile the summation can be replaced by an integral.
As mentioned before one measures the distortion and not the shear,
and equation~\ref{eqre} only works well when $\kappa$ is small.
Also substructure in the cluster core changes the average tangential
distortion (e.g., Hoekstra et al. 2000). To come around these
problems one has to choose an inner radius for the fit, such that
equation~\ref{eqre} is a good approximation.

If we assume that the tangential shear has been measured from a
distance $R_0$ out to $R$, the Einstein radius satisfies

\begin{equation}
r_E=\frac{2}{\ln(R/R_0)}\int_{R_0}^{R} dr~\gamma_T (r).
\end{equation}

Comparison with equation~\ref{eqmapq} shows that the best fit SIS model
is an aperture mass measurement where $Q(|\theta|)$ is given by

\begin{equation}
Q(|\theta|)=\left\{
\begin{array}{cl}
(\pi \theta \ln(R/R_0))^{-1} &, R_0\le|\theta|\le R\\
0 &, {\rm elsewhere}\\
\end{array}
\right. .
\end{equation}

The corresponding weight function for the surface density $U(\theta)$
is given by

\begin{equation}
U(|\theta|)=\frac{1}{\pi\ln(R/R_0)}
\left\{
\begin{array}{cl}
2/R_0-2/R	&, 0 \le |\theta| < R_0 \\
1/|\theta|-2/R 	&, R_0\le|\theta|\le R  \\
0		&, |\theta|>R 	     \\
\end{array}
\right. .
\end{equation}

The mass profiles of many clusters that have been studied to date are
consistent with a SIS model.  Thus the weight function suggested above
is close to the optimal choice. However, this choice for $U(|\theta|)$
is not a continuous function. A functional form which is continuous,
yet similar is suggested by S96, who discusses the use of aperture
mass statistics to find mass concentrations in deep images.

\subsection{$\zeta$-statistic}

Another statistic which has been widely used is the $\zeta$-statistic
(Fahlman et al. 1994):

\begin{equation}
\zeta(\theta_1,\theta_2)=\bar\kappa(<\theta_1)-
\bar\kappa(\theta_1<|\theta|<\theta_2).
\end{equation}

\noindent It gives the difference of the average dimensionless surface mass
density in a circular aperture of radius $\theta_1$ and that in an
annulus with $\theta_1<|\theta|<\theta_2$.  The corresponding weight
function $U(|\theta|)$ is given by:

\begin{equation}
U(|\theta|)=\left\{
\begin{array}{cl}
(\pi \theta_1^2)^{-1} &, |\theta| < \theta_1\\
-(\pi (\theta_2^2-\theta_1^2))^{-1} &, \theta_1<|\theta|<\theta_2\\
0 &, {\rm elsewhere}\\
\end{array}
\right. .
\end{equation}

\noindent This gives 

\begin{equation}
Q(|\theta|)=\left\{
\begin{array}{cl}
\theta_2^2(\pi \theta^2 (\theta_2^2-\theta_1^2))^{-1} &,  
\theta_1<|\theta|<\theta_2\\
0 &, {\rm elsewhere}\\
\end{array}
\right. .
\end{equation}

\noindent Thus the $\zeta$-statistic can be related to the tangential
shear through:

\begin{equation}
\zeta(\theta_1,\theta_2)=\frac{2\theta_2^2}{\theta_2^2-\theta_1^2}
\int_{\theta_1}^{\theta_2} d\ln\theta \langle\gamma_T\rangle(|\theta|).
\end{equation}

\begin{figure}
\begin{center}
\leavevmode
\hbox{%
\epsfxsize=8cm
\epsffile[35 180 300 670]{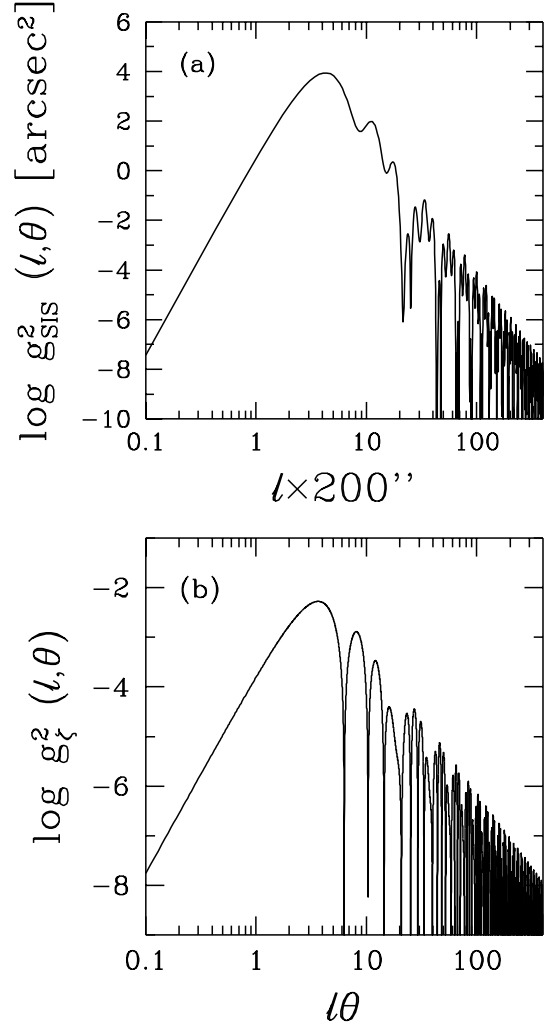}}
\begin{small}      
\caption{(a) The function $g_{\rm SIS}^2(l, \theta)$ as a function of
$l\theta$ for $\alpha=0.15$, and $\theta=200''$ (b)
$g_\zeta^2(l, \theta)$ for $\alpha=\frac{2}{3}$. Both curves show similar
behaviour, with $g_\zeta^2(l, \theta)$ peaking at a similar
values of $l\theta$.\label{filter}}
\end{small}
\end{center}
\end{figure}   

The SIS model fit has only a simple physical meaning when the cluster mass
profile is indeed isothermal. The interpretation of the $\zeta$-statistic
is independent of the mass distribution, although the statistic is
not optimal in terms of signal-to-noise ratio.

Under the assumption that one can estimate or minimize
$\bar\kappa(\theta_1<|\theta|<\theta_2)$, the average surface density
in the control annulus, one obtains a measurement of
$\bar\kappa(<\theta_1)$, and thus of the cluster mass. One way to do
so is to assume a model for the cluster surface mass density which is 
used to estimate the average surface density in the annulus 
(e.g. Hoekstra et al. 1998). This way the cluster mass estimate is
less dependent on the assumed model compared to the SIS model fit.

\section{Contribution by LSS}

Under the assumption that the cluster is the only lensing structure in
the field, the aperture mass $M_{\rm obs}$ provides a direct measure of the
cluster mass. However, the weak lensing signal is sensitive to all
matter along the line of sight, and other structures, which we will
refer to as large scale structure, will introduce an additional
signal:

\begin{equation}
M_{\rm obs}(\theta)=M_{\rm cluster}(\theta)+M_{\rm LSS}(\theta).
\end{equation}

The aperture mass statistic measures a density contrast, and the
expectation value for the contribution by uncorrelated large scale structure
$\langle M_{\rm LSS}\rangle=0$ (see also SWJK). Thus on average,
distant structures do not introduce a bias in cluster mass estimates,
but an uncertainty in the cluster mass of \map. 
As a result the effect of distant large scale structure can be treated
as an additional (statistical) source of error. Its relative
importance can be decreased by averaging the results for several
clusters, i.e.  measuring their average mass. We note that on certain
scales $M_{\rm LSS}$ might be distributed skewly.

The effect of `local' (or correlated) large scale structure has been
studied using N-body simulations (e.g., Cen 1997; Metzler et
al. 1999).  In this case either projection effects, or filaments
connected to the cluster bias the weak lensing mass estimates towards
somewhat higher values. This becomes particularly important when one
tries to estimate the enclosed mass within several Mpc's from the
cluster centre (Cen 1997).

SWJK have shown how one can relate the aperture mass statistic to the
power spectrum of density fluctuations. Measuring \map~for a number of
random fields allows one to constrain the projected power spectrum.
In this paper the problem is reversed: given the power spectrum, what
is the resulting \map? A detailed discussion about the use of aperture
mass statistics and cosmic shear can be found in SWJK and Bartelmann
\& Schneider (1999). 

SWJK showed that the effect of large scale structure on the variance
in the aperture mass statistic is given by:

\begin{equation}
\langle M_{\rm LSS}^2\rangle(\theta)= 2\pi\int_0^\infty dl~l P_\kappa(l) 
g(l,~\theta)^2, \label{eqmap}
\end{equation}

\noindent where the effective projected power spectrum $P_\kappa(l)$ is 
given by:

\begin{equation}
P_\kappa(l)=\frac{9 H_0^4 \Omega_m^2}{4 c^4}
\int_0^{w_H} dw \left(\frac{\bar W(w)}{a(w)}\right)^2 
P_\delta\left(\frac{l}{f_K(w)};w\right)
\end{equation}

\noindent Here $w$ is the radial coordinate, $a(w)$ the cosmic scale factor,
and $f_K(w)$ the comoving angular diameter distance. $\bar W(w)$ is the
source averaged ratio of angular diameter distances $D_{ls}/D_{s}$ for
a redshift distribution of sources $p_w(w)$:

\begin{equation}
\bar W(w)=\int_w^{w_H} dw' p_w(w')\frac{f_K(w'-w)}{f_K(w')}.
\end{equation}

\noindent The function $g(l,~\theta)$ in equation~\ref{eqmap} depends on
the filter function $U(|\theta|)$, and is given by:

\begin{equation}
g(l,~\theta)=\int_0^\theta d\phi~\phi U(\phi)J_0(l\phi).
\end{equation}

As has been demonstrated in Jain \& Seljak (1997) and SWJK it is
crucial to use the non-linear power spectrum in equation~20. This
power spectrum can be derived from the linearly evolved cosmological
power spectrum following the prescriptions of Hamilton et al. (1991),
and Peacock \& Dodds (1996).

\subsection{Filter functions}

The form of the function $g^2(l,~\theta)$, which probes the projected
power spectrum, depends on the choice of $U(|\theta|)$. For the
SIS model fit we take $R_0=\alpha R$, which yields

$$g_{\rm SIS}(l,~\theta) = \frac{-\theta}{\pi\ln(\alpha)l \theta} 
\left[2\left\{J_1(\alpha l\theta)-J_1(l\theta)\right\} +  
\int\limits_{\alpha l\theta}^{l\theta} ds J_0(s)\right].$$
\begin{equation}
~
\end{equation}

\noindent We will use a typical value for $\alpha=0.15$ (e.g.,
Hoekstra et al. 1998). Larger values are used when much substructure
is found in the cluster core (e.g., Hoekstra et al. 2000). The
resulting $g_{\rm SIS}^2(l,~\theta)$ is presented in
figure~\ref{filter}a.

\noindent For the $\zeta$-statistic we find 

\begin{equation}
g_\zeta(l~,\theta)=\frac{J_1(\alpha l \theta)-\alpha J_1(l \theta)}
{\pi \alpha l \theta (1-\alpha^2)},
\end{equation}

\noindent where we have taken the inner radius
$\theta_1=\alpha\theta$.  A typical value for $\alpha$ used in weak
lensing mass estimates of clusters is $\frac{2}{3}$, i.e., the control
annulus runs from the boundary of the region in which we want to
determine the mass out to 1.5 times this radius. Throughout this paper
we will use this value for $\alpha$. In figure~\ref{filter}b the
filter function $g_\zeta^2(l,~\theta)$ is shown (taking
$\alpha=\frac{2}{3}$).

\begin{figure}
\begin{center}
\leavevmode
\hbox{%
\epsfxsize=8cm
\epsffile{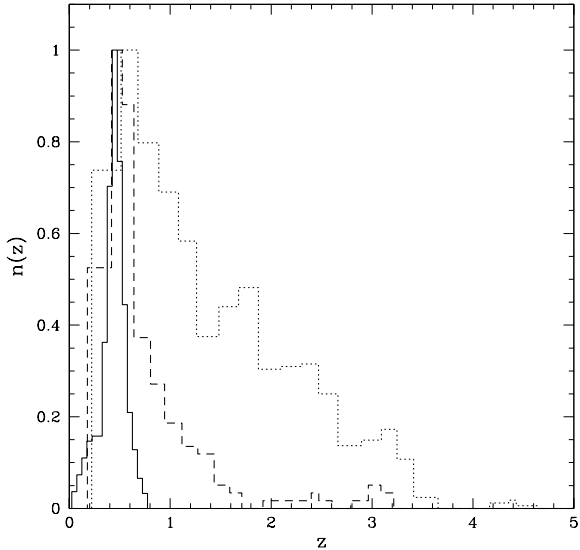}}
\begin{small}      
\caption{Photometric redshift distributions from the HDF north and
south.  The counts have been normalized to a peak value of unity. The
solid line is for galaxies with $20<R<22$, the dashed line is for
$20<R<24$, and the dotted line for $20<R<26$. These distributions are
used to compute the lensing signals and the contribution of large
scale structure.
\label{zdist}}
\end{small}
\end{center}
\end{figure}   

\subsection{Photometric redshifts}

The projected power spectrum $P_\kappa(k)$ depends on the redshift
distribution of the sources, and consequently we need to know, or
assume, a redshift distribution for the sources. We will use the
photometric redshift distributions inferred from the northern and
southern Hubble Deep Fields (Fern{\'a}ndez-Soto, Lanzetta, Yahil
1999; Chen et al. 1998), which generally work well (e.g.,
Hoekstra, Franx, \& Kuijken 2000). There is, however, some field
to field variation in the redshift distribution the sources, but
we assume that the distributions obtained from the HDFs are
representative for the whole sky. We note that most of the lensing
by large scale structure is caused by structures at intermediate 
redshifts, and as a result the signal induced by the LSS is
fairly robust against variations in the source redshift distribution.

Using the colours of the galaxies in the HDFs we have computed their
$R$ magnitudes, which allows a direct comparison with ground based
observations. We use galaxies fainter than $R=20$. The resulting
photometric redshift distributions for different limiting magnitudes
are presented in figure~\ref{zdist}.

\begin{figure}
\begin{center}
\leavevmode
\hbox{%
\epsfxsize=8cm
\epsffile[15 170 300 610]{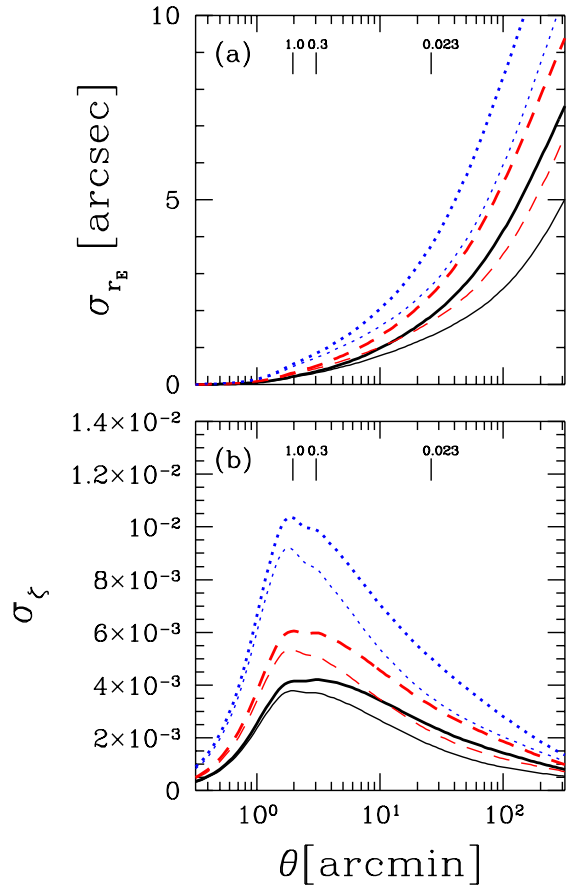}}
\begin{small}      
\caption{(a) Scatter in the estimate of the Einstein radius $r_E$ 
due to large scale structure for the SCDM model (thick lines)
and the OCDM model (thin lines). We use three different redshift
distributions for each model. These are based on photometric redshifts
inferred from both HDFs taking different limiting magnitudes.  The
solid line corresponds to $20<R<22$, the dashed line to $20<R<24$, and
the dotted line to $20<R<26$. We have indicated a radius of
$1~h_{50}^{-1}$~Mpc for clusters at various redshifts. Panel (b)
shows the results for the $\zeta$-statistic as a function of aperture
size.
\label{mapscale}}
\end{small}
\end{center}
\end{figure}   

\begin{table}[b]
\begin{center}
\begin{tabular}{lcccc}
\hline
\hline
   		& $h$ & $\Omega_m$ & $\Omega_\Lambda$ & $\sigma_8$ \\
SCDM    	& 0.5 & 1.0 	   & 0.0	      & 0.5  \\
OCDM    	& 0.7 & 0.3	   & 0.0	      & 0.85 \\
$\Lambda$CDM    & 0.7 & 0.3	   & 0.7	      & 0.90 \\
\hline
\hline
\end{tabular}
\begin{small}
\caption{Parameters of the cosmological models considered here. All
three models are cluster normalized.\label{cosmotab}}
\end{small}     
\end{center}
\end{table}

\begin{figure*}
\begin{center}
\leavevmode
\hbox{%
\epsfxsize=8cm
\epsffile[160 175 570 700]{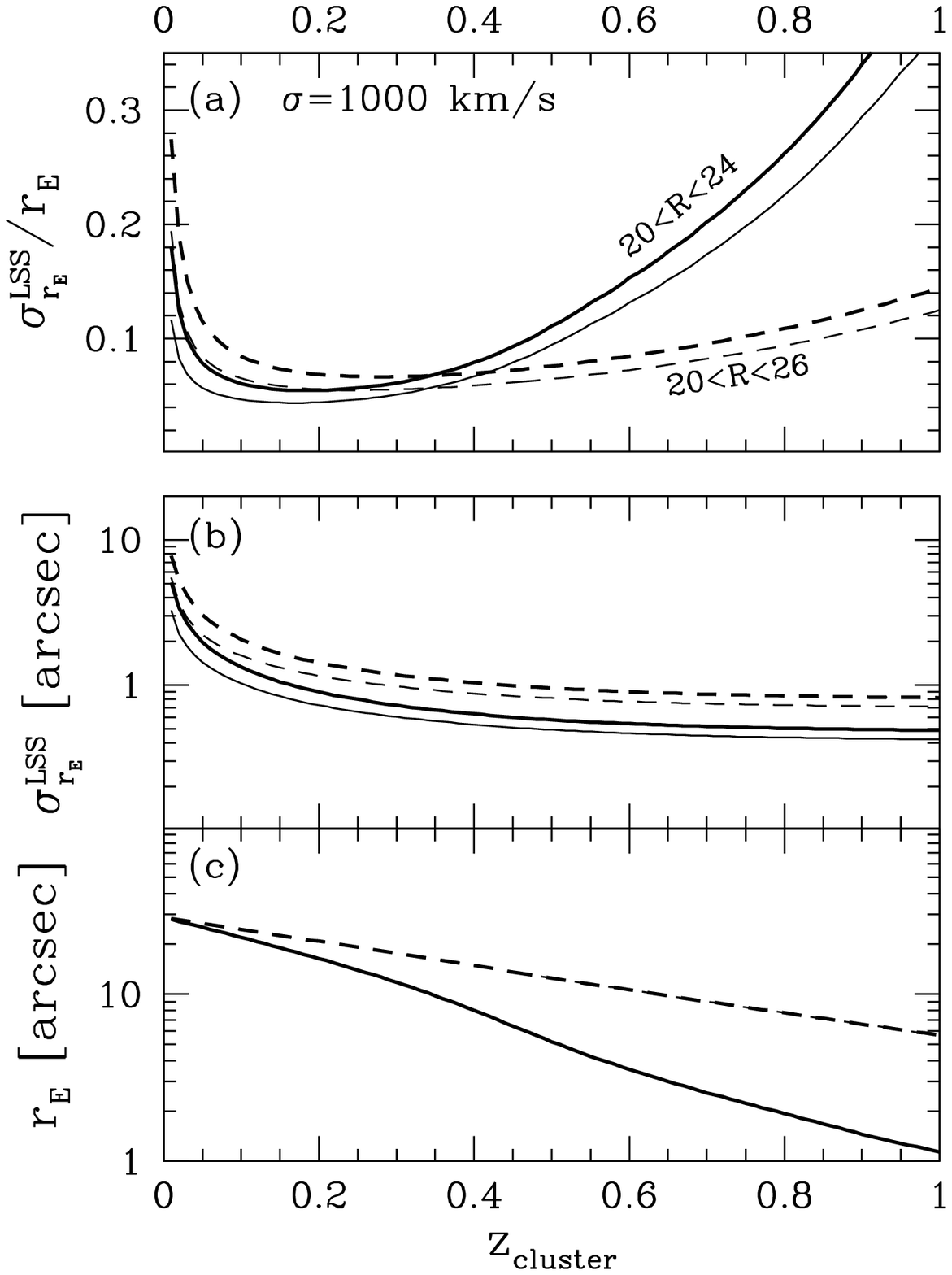}
\epsfxsize=8cm
\epsffile[160 175 570 700]{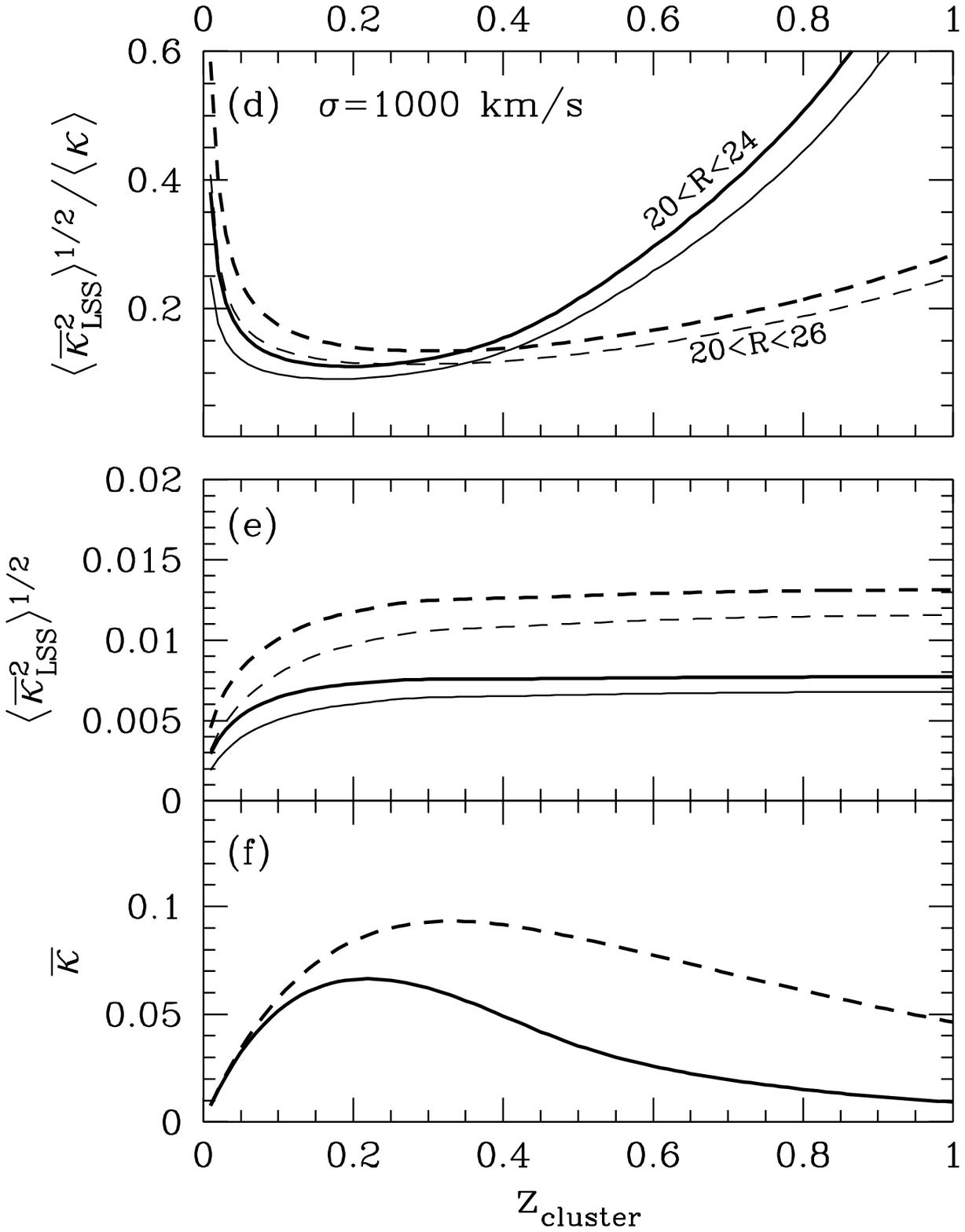}}
\begin{small}      
\caption{(a) Relative importance of lensing by distant large scale
structure on the measurement of the Einstein radius $r_E$ as a
function of cluster redshift, when the shear is measured in an
annulus from 0.225 to $1.5~h_{50}^{-1}$~Mpc. (b) the scatter in the
measurement of $r_E$ caused by large scale structure. (c) The observed
Einstein radius as a function of cluster redshift. (d) same as panel
(a), but now when the $\zeta-$statistic is used to estimate
$\bar\kappa(<1~h_{50}^{-1}~{\rm Mpc})$. We have assumed that the
lensing signal is measured out to $1.5~h_{50}^{-1}$ Mpc from the
cluster centre, and that the average surface density
$\bar\kappa(1<r<1.5~h_{50}^{-1}~ {\rm Mpc})$ is determined from a SIS
model fit to the tangential shear.  (e) scatter in $\bar\kappa$ because
of large scale structure. (f) $\bar\kappa(<1~h_{50}^{-1}~{\rm Mpc})$
as a function of cluster redshift. The thick lines
correspond to the SCDM model, and the thin lines to the open
model. The values for either $r_E$ and $\bar\kappa$ turn out to be
almost identical for these two cosmologies.\label{ratio}}
\end{small}
\end{center}
\end{figure*}   

\subsection{Aperture statistics}

Given the photometric redshift distribution discussed above we
calculate the contribution of the LSS to the measurements of $r_E$ and
the $\zeta$-statistic as a function of aperture size. We have
considered three different cosmologies, which are listed in
table~\ref{cosmotab}.
                 
Figure~\ref{mapscale}a shows the scatter in the value of the Einstein
radius $r_E$ as a function of aperture size for the SCDM (thick lines)
and the OCDM (thin lines) models. For clarity, we limit the number of
curves shown in figure~\ref{mapscale} and omit the results for the
$\Lambda$CDM model. For this model the scatter is similar to the OCDM model,
with a somewhat higher amplitude on large scales and a somewhat lower
amplitude on arcminute scales. The scatter in the value of $r_E$
increases for larger aperture sizes. The value of $r_E$ is independent
of aperture size, and as a result the relative contribution of large
scale structure to the error budget increases with increasing aperture
size. Also the scatter is larger for fainter limiting magnitudes,
i.e. higher source redshifts.

Figure~\ref{mapscale}b shows the scatter for the $\zeta$-statistic.
Because both $\langle\zeta^2_{LSS}\rangle^{1/2}$ and
$\bar\kappa(<\theta)$ decrease with increasing aperture size the
interpretation of figure~\ref{mapscale}b is not as straightforward as
is the case for figure~\ref{mapscale}a. The effect of aperture size
on the accuracy of weak lensing mass estimates is discussed in detail
in section~4.3.

Comparison of the results for the SCDM and OCDM cosmologies indicates
that the differences are fairly small. We found that the $\Lambda$CDM
cosmology gives similar estimates. Although we consider these
different cosmologies throughout the paper (and plot the results for
the OCDM and SCDM models) we note that the results hardly depend on
the assumed cosmology.

\section{Results}

We examine the uncertainty in the weak lensing mass estimate caused by
uncorrelated large scale structure. The results are derived for a
cluster with a SIS mass profile, and a velocity dispersion of
$\sigma=1000$ km/s, typical for rich clusters that have been studied
in the past. We study the dependence on cluster redshift, where we
discuss the results for low redshift clusters in more detail.
We also examine how the accuracy of the mass measurements depends
on aperture size.

\subsection{Dependence on cluster redshift}

First we examine how the mass estimates for clusters at different
redshifts are affected when the lensing signal is measured out to
a radius of $1.5~h_{50}^{-1}$ Mpc.

For the SIS model we use the measurements in an annulus from 0.225 to
$1.5~h_{50}^{-1}$~Mpc ($R=1.5~h_{50}^{-1}$ Mpc, $\alpha=0.15$).  In
the case of the $\zeta-$statistic we use an inner radius of
$\theta_1=1~h_{50}^{-1}$ Mpc and an outer radius of
$\theta_2=1.5~h_{50}^{-1}$ Mpc $(\alpha=\frac{2}{3})$. The value of the
$\zeta$-statistic only provides a lower limit to the cluster mass, and
one way to circumvent this is to estimate the average surface density
in the annulus around the aperture, for instance by using the results
from a model fit. However, this model fit is also affected by
large scale structure. In general the region where the model is fitted
to the data largely coincides with the annulus used to compute the
value of the $\zeta$-statistic. Thus if the value of the
$\zeta$-statistic is increased by additional structures, the mass from
the model will also be higher. Although the uncertainty in the model
parameters and the $\zeta$-statistic are not completely correlated, we
will assume they are. This gives a fair upper limit on the uncertainty
in the estimate of $\bar\kappa$

\begin{equation}
\sigma^{\rm LSS}_{\bar\kappa}=\sigma^{\rm LSS}_\zeta +
\frac{\sigma^{\rm LSS}_{r_E}}{(1+\alpha) \theta_2}.
\end{equation}

The results are presented in figure~\ref{ratio}. Panels~c and~f
show respectively the value of the Einstein radius and $\bar\kappa$
as a function of cluster redshift. We show the results for the
SCDM (thick lines) and the OCDM (thin lines) cosmologies. The solid
lines are for a limiting magnitude of $R=24$ and the dashed lines
for $R=26$. Panels~b and~e show the scatter in the aperture mass
statistics introduced by the large scale structure. Finally the
ratios of the scatter and the lensing signal are presented in
figure~a and~d.

\begin{figure}
\begin{center}
\leavevmode
\hbox{%
\epsfxsize=8cm
\epsffile[35 170 350 690]{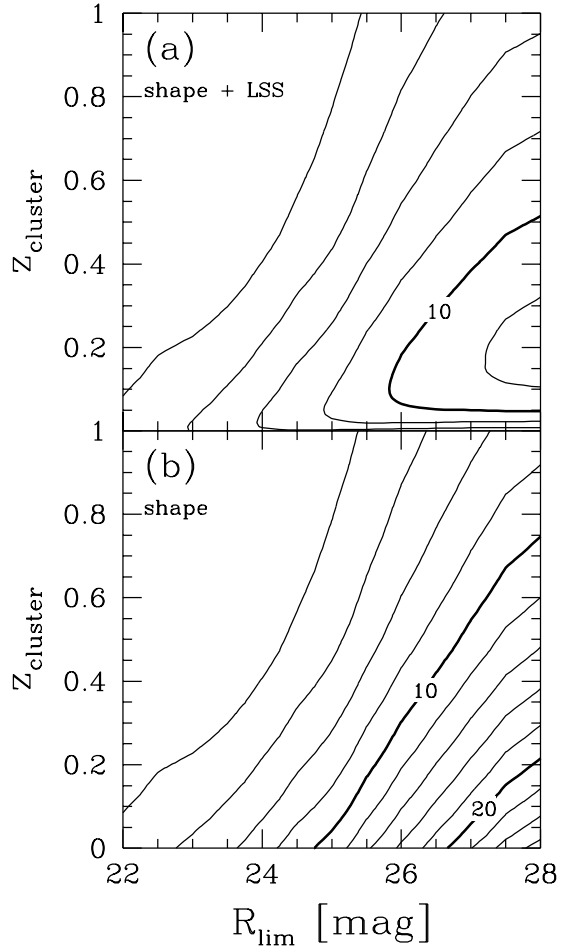}}
\begin{small}      
\caption{(a) Contour plot of the signal-to-noise ratio for a SIS model fit
to the tangential shear of a cluster of $\sigma=1000$ km/s as a function
of limiting magnitude $R_{\rm lim}$ and cluster redshift $z_c$.
The noise includes the contribution of the intrinsic shapes of the
sources and the contribution by large large scale structure for
the OCDM cosmology.
(b) Contour plot of the signal-to-noise ratio when the noise
caused by large scale structure is ignored. The interval between
adjacent contours is 2.
\label{contsn}}
\end{small}
\end{center}
\end{figure}  

For clusters at low redshifts $(z<0.2)$ the relative importance
of large scale structure for the $\Lambda$CDM model is similar
to that of the SCDM model. For clusters at high redshift, or
deep observations, the results for the $\Lambda$CDM are close
to the OCDM results.

Comparison of figures~\ref{ratio}a and~d shows that the relative
importance of the distant large scale structure is smaller in the case
of the SIS model fit. For this mass estimator the uncertainty in the
cluster mass estimate is about 6\% for rich clusters at intermediate
redshifts. The uncertainty in the mass is large for high redshift
clusters when the limiting magnitude of the sources is rather bright
(solid lines). However, in practice, to detect a significant lensing
signal of a high redshift cluster deep observations are needed to
ensure a sufficiently high number density of sources. As a result the
importance of large scale structure is minimized (dashed lines).  In
the case of nearby clusters, the presence of structure along the line
of sight results in a relatively large uncertainty in the cluster mass
(see section~4.2).

The uncertainty introduced by large scale structure is much larger
when the $\zeta$-statistic is used. This is not surprising as one
only uses the shear measurements at relatively large radii, where
the cluster mass surface density has become fairly low. In addition
to this annulus, the SIS model fit uses data at much smaller radii,
where the cluster signal is much higher.

The noise in weak lensing analyses is generally dominated by the
intrinsic shapes of the source galaxies, and not by the noise in the
large scale structure. Both sources of error depend differently on the
limiting magnitude of the background galaxies, and as a result the
total noise is a rather complex function of cluster redshift and
limiting magnitude, as is demonstrated in figure~\ref{contsn}a for
the OCDM cosmology.

Figure~\ref{contsn}b shows the the signal-to-noise ratio for a SIS
model fit to the tangential shear of a cluster of $\sigma=1000$ km/s,
when the contribution of large scale structure is ignored. It should
be noted here that we assume that the shapes of the sources have been
measured accurately. In reality the shapes of very faint galaxies are
difficult to determine. From figure~\ref{contsn}b one would expect to
be able to measure accurate masses for low redshift clusters.
However, the results change significantly when we include the
contribution from large scale structure as is shown in
figure~\ref{contsn}a. The change in signal-to-noise ratio is
small for high redshift clusters, where the shapes are still the
major contributor to the noise. The true signal-to-noise ratio
is lowered significantly in the case of deep observation of
clusters at intermediate redshifts, but the largest difference
is seen for clusters at low redshifts.

The signal-to-noise ratios for the $\Lambda$CDM model are intermediate
to those found for the OCDM and SCDM models, resembling the OCDM
better for high redshift clusters. However, at high redshifts the
differences between the various models become less important (they are
fairly small anyway), because the noise is dominated by the intrinsic
shapes of the sources.

\subsection{Low redshift clusters}

Figure~\ref{contsn}a demonstrates that large scale structure introduces a
large uncertainty in the mass measurement of low redshift clusters.
To date not many low redshift clusters have been studied through weak
lensing. This is because the amplitude of the lensing signal peaks
at intermediate redshifts, and the signal is extremely small for
nearby clusters. Furthermore their angular sizes were too large for
the CCD cameras. Panoramic CCD cameras have become available recently,
and a number of low redshift clusters have been imaged for
weak lensing purposes. For instance Joffre et al. (2000) have
analysed the mass distribution of Abell 3667 $(z=0.055)$.

Although the lensing signal of low redshift clusters is small, there
are several reasons to study these systems. First of all, they can, or
have been studied extensively by other techniques. Also, the
uncertainty in the redshift distribution of the sources is almost
non-existent, and one can use relatively large background
galaxies. 

In this section we examine how the weak lensing mass measurement of
the Coma cluster $(z=0.023)$ is affected by structures at higher
redshifts. A weak lensing analysis of the most nearby rich cluster of
galaxies is challenging, but could provide important insights in the
cluster dynamics.  We assume a velocity dispersion of 1000~km/s for
the cluster (e.g., Colles \& Dunn 1996).

\begin{figure}
\begin{center}
\leavevmode
\hbox{%
\epsfxsize=\hsize
\epsffile{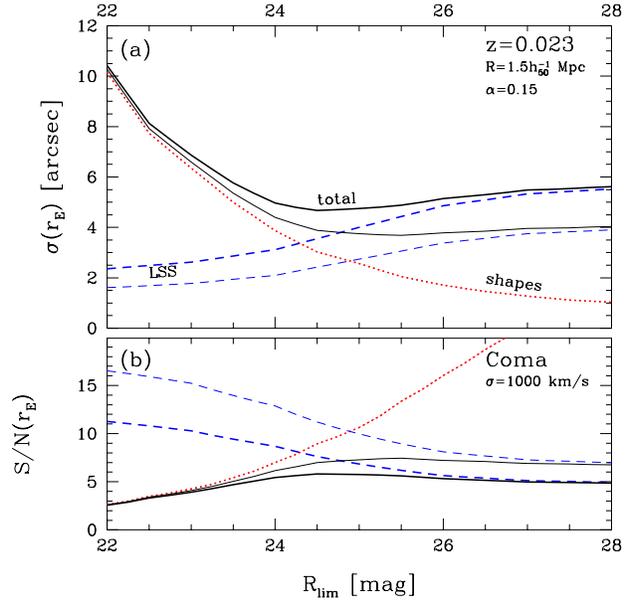}}
\begin{small}      
\caption{(a) Sources of noise in the measurement of the Einstein
radius $r_E$ from a SIS model fit.  The figure shows the noise caused
by the large scale structure (dashed lines) and the intrinsic shapes
of the sources (dotted lines) as a function of limiting magnitude
$R_{\rm lim}$ of the sources. The thick lines correspond to the SCDM
model and the thin lines to the OCDM cosmology.  The results for the
$\Lambda$CDM model are similar to the SCDM model. (b) Expected
signal-to-noise ratio as a function of limiting magnitude of the
sources for the Coma cluster. For limiting magnitudes $R>24$
the signal-to-noise ratio of the measurement of the Einstein radius
no longer increases, because of the noise from the large scale 
structure. The results for the $\Lambda$CDM model lie between the
drawn curves.
\label{noise_SIS}}
\end{small}
\end{center}
\end{figure}  

In figure~\ref{noise_SIS}a we plot the sources of noise in the
measurement of the Einstein radius as a function of limiting magnitude
of the sources. The figure shows the contribution from the large scale
structure (dashed lines) and the intrinsic shapes of the sources
(dotted lines). For the intrinsic shapes of the sources we have
assumed $\langle \gamma_T^2\rangle^{1/2}=0.25$, similar to
what is observed in deep HST images (e.g., Hoekstra et al. 2000)
Reaching the depth of the HDFs one would expect to obtain a
signal-to-noise ratio of more than 20. However,
figure~\ref{noise_SIS}a demonstrates that the total uncertainty in the
measurement of the Einstein radius reaches a minimum value for a
limiting $R=24$.

Figure~\ref{noise_SIS}b shows the expected signal-to-noise ratio as a
function of limiting magnitude, and it demonstrates that observations
with $R_{\rm lim}>24.5$ no longer improve the signal-to-noise ratio of
the mass measurement. The maximum signal-to-noise ratio that can be
reached by fitting a SIS model to the observed tangential shear around
the Coma cluster is $\sim 5$ for the SCDM model, and $\sim 7$ for the
OCDM model. The results for the $\Lambda$CDM model lie between the
curves for the other two models.

\begin{figure}
\begin{center}
\leavevmode
\hbox{%
\epsfxsize=\hsize
\epsffile{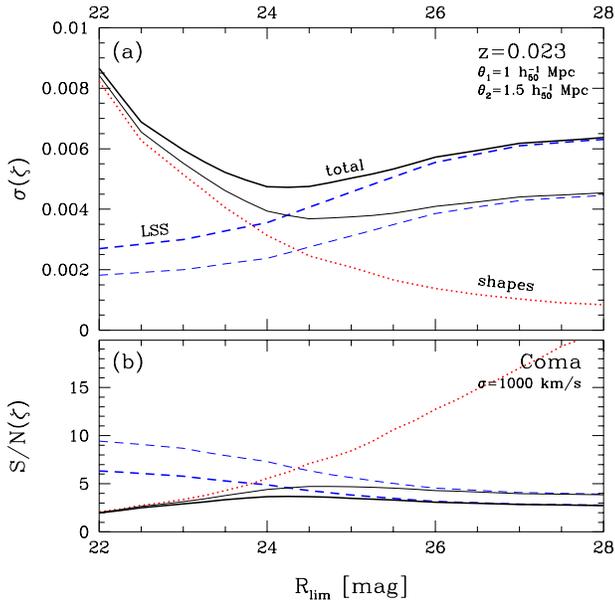}}
\begin{small}      
\caption{As figure~\ref{noise_SIS} but now for the $\zeta-$statistic.
The noise introduced by the large scale structure results in a 
constant signal-to-noise ratio for limiting magnitudes $R>25$.
Note that the achieved signal-to-noise ratio for the $\zeta$-statistic
is lower in comparison to the SIS model fit.
\label{noise_zeta}}
\end{small}
\end{center}
\end{figure}  

In figure~\ref{noise_zeta} we show how the accuracy of the mass
measurement is affected by large scale structure when the
$\zeta$-statistic is used. Comparison with figure~\ref{noise_SIS}
shows that the SIS model fit yields a better signal-to-noise ratio. 

\subsection{Large apertures}

The intrinsic shapes of the sources result in a statistical uncertainty
in the measurement of the Einstein radius. If the SIS model is fitted
to the measurements in an annulus from $R_0$ out to $R$, the error
in $r_E$ is given by

\begin{equation}
\sigma_{r_E}=\sqrt{\frac{2}{\pi \bar n \ln (R/R_0)}} \sigma_{\rm gal},
\label{noisere}
\end{equation}

\noindent where $\bar n$ is the number density of sources, and 
$\sigma_{\rm gal}$ is the uncertainty in the measurement of the
tangential distortion for a single galaxy. 

From equation~\ref{noisere} we find that this contribution to the
error budget decreases slowly for increasing aperture sizes. However,
figure~\ref{mapscale}a shows that the noise contribution from large
scale structure increases rather rapidly with aperture size.

\begin{figure}
\begin{center}
\leavevmode
\hbox{%
\epsfxsize=\hsize
\epsffile{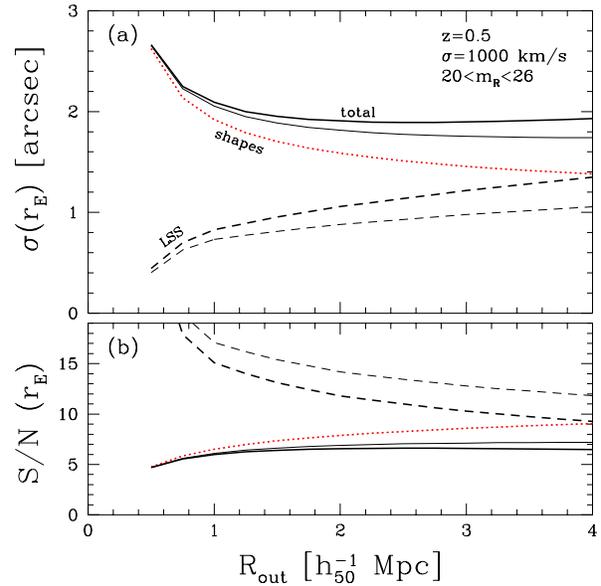}}
\begin{small}      
\caption{(a) Error on the measurement of the Einstein radius $r_E$ as
a function of aperture size, because of the large scale structure
(dashed lines) and the instrinsic shapes of the sources (dotted
lines). The SIS model is fitted to the tangential shear in an annulus
ranging from $R_0=0.225~h_{50}^{-1}$ Mpc out to $R_{\rm out}$.  The
thick lines give the results for the SCDM model, whereas the thin
lines correspond to the OCDM model.  (b) The expected signal-to-noise
ratio as a function of aperture size. The curves for the $\Lambda$CDM
model are comparable to the results for the OCDM model. The signal-to-noise
ratio does not improve for aperture sizes $R_{\rm out}>1~h_{50}^{-1}$ Mpc.
\label{noise_app_SIS}}
\end{small}
\end{center}
\end{figure}  

Figure~\ref{noise_app_SIS}a shows how the contributions of the
intrinsic shapes of the sources, and the large scale structure depend
on the aperture size. The results are for a cluster at a redshift
$z=0.5$, and the lensing signal has been measured from
$R_0=0.225~h_{50}^{-1}$ Mpc out to a radius $R_{\rm out}$. The
uncertainty in $r_E$ caused by the intrinsic shapes of the sources
decreases slowly with aperture size (dotted curve), whereas the
contribution by the distant large scale structure increases (dashed
line). Figure~\ref{noise_app_SIS}b shows the resulting signal-to-noise
ratio. The signal-to-noise ratio is virtually constant for outer radii
larger than $\sim 1~h_{50}^{-1}$ Mpc. Measuring the lensing signal in
larger apertures does not improve the accuracy of the cluster mass.
One has also to keep in mind that we have assumed that the mass profile is
well described by a SIS model, which might not be true in the outer
parts of clusters. 

For clusters at higher redshifts the curves look similar, but the
maximum signal-to-noise ratio is reached at somewhat larger (physical)
aperture sizes. The noise caused by the large scale structure
dominates the error for low redshift clusters, and for these systems,
the signal-to-noise ratio decreases slowly with increasing aperture
size.

So far we have used an inner radius for the SIS model fit of 
$R_0=0.225~h_{50}^{-1}$ Mpc, but one cannot always fit the model at
such small distances from the centre. For instance in the case
of the $z=0.83$ cluster MS~1054-03 Hoekstra et al. (2000) showed that
the mass distribution in the centre is rather complex and the substructure
lowers the azimuthally averaged tangential shear. Consequently they
fit the SIS model from $\sim 0.7$ out to $\sim 2~h_{50}^{-1}$ Mpc,
which corresponds to $\alpha=0.35$. 

We also have examined how the accuracy of the mass estimate changes when
the SIS model is fitted to the data within a given aperture, while the 
inner radius is increased. In this situation both the noise caused by
the shapes of the sources and the contribution by the large scale
structure increase. Thus small values for the inner radius improve
the signal-to-noise of the mass measurement, but one should keep in
mind that substructure in the cluster centre complicates a safe
choice for the inner radius.

For the $\zeta$-statistic the noise contribution caused by the shapes
of the galaxies is given by

\begin{equation}
\sigma_\zeta=\sqrt{\frac{1}{\pi(1-\alpha^2)\theta_1^2 \bar n}}
\sigma_{\rm gal}.
\end{equation}

However, we want to estimate $\bar\kappa(<1~h_{50}^{-1}{\rm Mpc})$,
and thus we have to add the average surface density in the annulus
from 1 to $R_{\rm out}$ $h_{50}^{-1}$ Mpc. As before we estimate
$\bar\kappa(1<r<R_{\rm out})$ from a SIS model fit to the tangential
shear. The errors in the $\zeta-$statistic and the Einstein radius
from the model fit are combined according to equation~25, because
the errors are correlated. We note that the size of the correction
depends on the assumed mass profile, but for increasing values of
$R_{\rm out}$ the correction becomes smaller (i.e., 
$\bar\kappa(1<r<R_{\rm out})$ becomes smaller).

\begin{figure}
\begin{center}
\leavevmode
\hbox{%
\epsfxsize=\hsize
\epsffile{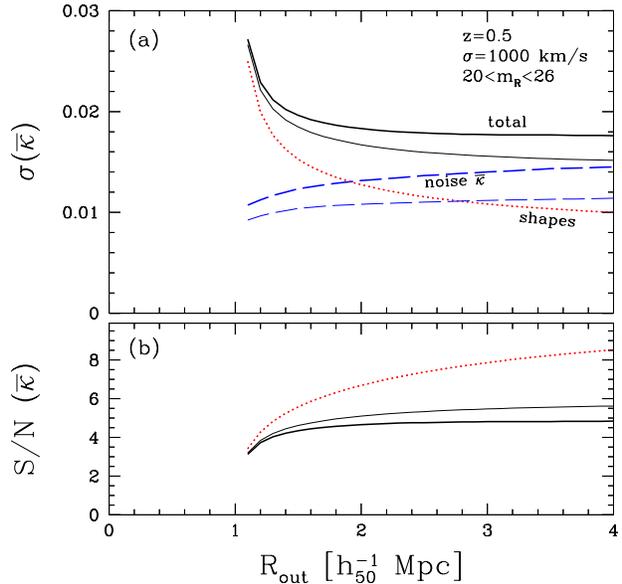}}
\begin{small}      
\caption{As figure~\ref{noise_app_SIS}, but now we show the measurement
error on $\bar\kappa(<1~h_{50}^{-1}{\rm Mpc})$ as a function of
$R_{\rm out}$ when the $\zeta-$statistic is used to estimate the mass.
The average surface density in the annulus from 1 to $R_{\rm out}$
$h_{50}^{-1}$ Mpc is estimated from a SIS model fit to the
observations.  The thick lines give the results for the SCDM model,
whereas the thin lines correspond to the OCDM model. (b) The expected
signal-to-noise ratio of $\bar\kappa(<1~h_{50}^{-1} {\rm Mpc})$ as a
function of aperture size. The signal-to-noise ratio improves with
increasing $R_{\rm out}$. If the cluster mass profile is well described
by a SIS model, the model fit gives the better results.
\label{noise_app_zeta}}
\end{small}
\end{center}
\end{figure}

The resulting noise estimates are presented in
figure~\ref{noise_app_zeta}a.  The corresponding signal-to-noise ratio
of $\bar\kappa(<1~h_{50}^{-1}{\rm Mpc})$ for a cluster at a redshift
$z=0.5$ is presented in figure~\ref{noise_app_zeta}b. It shows that
using a larger $R_{\rm out}$ improves the accuracy in the measurement
of $\bar\kappa(<1~h_{50}^{-1}{\rm Mpc})$.

A similar calculation shows that the contribution of the large
scale structure becomes more important when one tries to measure
the average surface density in larger apertures. For instance
the uncertainty in $\bar\kappa(<2~h_{50}^{-1}{\rm Mpc})$ almost
doubles compared to $\bar\kappa(<1~h_{50}^{-1}{\rm Mpc})$.

Attempts to measure the average cluster mass profile out to large
projected distances from the cluster centre will also be affected by
large scale structure, given the fact that it is already difficult
to measure the total mass. Unfortunately it will be difficult to
quantify the effect of distant large scale structure on such
measurements.

\section{Conclusions}

The observed weak lensing signal is sensitive to all matter
along the line of sight, and as a result this additional matter acts
as a source of uncertainty in mass estimates of clusters of
galaxies. In this paper we have investigated the effect of distant
large scale structure, uncorrelated with the cluster mass
distribution. It can be treated as an additional source of noise.

Given a cosmological model it is possible to compute the statistical
uncertainty in the weak lensing mass measurement caused by the distant
large scale structure, and quantify its contribution to the total
error budget.  We have considered two methods to estimate the mass
from a weak lensing analysis. One way is to fit a SIS model to the
observed lensing signal. The other method uses the $\zeta$-statistic
(e.g., Fahlman 1994) to estimate the mass. Both methods give similar
results.

We find that the importance of matter along the line sight is fairly
small for rich clusters ($\sigma=1000$ km/s) at intermediate
redshifts, provided that the bulk of the sources are at high redshift
compared to the cluster. If the lensing signal is measured out to
$1.5~h_{50}^{-1}$ Mpc from the cluster centre, the typical $1 \sigma$
relative uncertainty in the mass cause by the large scale structure is
about $6\%$.

In certain situations the effect of large scale structure is more
important. For nearby clusters, such as the Coma cluster, background
structures introduce a considerable uncertainty in the
mass. Considering only the noise caused by the shapes of the sources
used in the weak lensing analysis one expects to achieve very accurate weak
lensing mass measurements of nearby clusters, but we find that the
large scale structure limits the maximum achievable signal-to-noise
ratio to $\sim 7$, irrespective of the depth of the observations. 

We also have examined how the accuracy of the weak lensing mass
estimate changes when the lensing signal is measured out to large
projected distances from the cluster centre. In the case of the SIS
model fit we find that the signal-to-noise ratio does not increase
significantly when one includes measurements at radii larger than
$\sim 1~h_{50}^{-1}$ Mpc. When the $\zeta$-statistic is used to
estimate the mass within a $1~h_{50}^{-1}$ Mpc radius aperture the
signal-to-noise ratio does increase with $R_{\rm out}$. If the
cluster mass profile well described by a SIS model, the model
fit results in a better signal-to-noise ratio.

The effect of large scale structure is important in the outskirts of
clusters, where the cluster signal itself is low.  In this paper we
have quantified the uncertainty in the measurement of the cluster
mass, but distant large scale structure will also introduce noise in
studies that try to constrain the mass profiles of clusters at large
projected distances.  Furthermore studies of the 3-D structure of
clusters by combining weak lensing, X-ray, and S-Z data (e.g., Zaroubi
et al. 1998) will be affected by uncorrelated large scale structure.

Photometric redshifts of the sources can be used to decrease the
effect of large scale structure, either by modeling the contribution
of large scale structure to the lensing signal, or by selecting
sources in an optimal redshift range. This can be only done at the
expense of a large amount of additional data. Because the effect of
large scale structure can be treated as an additional (statistical)
source of noise. Thus an alternative procedure is to average the
results of an ensemble of clusters, thus decreasing the noise on the
(average) cluster mass estimate.

\begin{acknowledgements}
We thank the referee for several useful comments which improved the
paper. It is a pleasure to thank Marijn Franx and Konrad Kuijken for
useful discussions and comments to improve the paper. Also the
comments from Ludo van Waerbeke are acknowledged. Peter Schneider made
available his algorithms, which were very helpful.
\end{acknowledgements}

\end{document}